\documentclass[modern]{aastex62}

\hypersetup{linkcolor=black,citecolor=black,filecolor=cyan,urlcolor=magenta}

\submitjournal{ApJL}

\shorttitle{Infrared Detection of CS in AFGL 2591}
\shortauthors{Barr et al.}

\begin{document}

\title{Infrared Detection of Abundant CS in the Hot Core AFGL 2591\\ at High Spectral Resolution with SOFIA/EXES \protect\footnote{Observations made at the IRTF and SOFIA} }

\correspondingauthor{Andrew G. Barr}
\email{barr@strw.leidenuniv.nl}

\author[0000-0003-4909-2770]{Andrew G. Barr}
\affiliation{Leiden University \\
Niels Bohrweg 2, 2333 CA Leiden, \\
The Netherlands}

\author[0000-0001-9344-0096]{Adwin Boogert}
\altaffiliation{Staff Astronomer at the
 Infrared Telescope Facility, which is operated by the University of
 Hawaii under contract NNH14CK55B with the National Aeronautics and
 Space Administration.}
\affiliation{Institute for Astronomy \\
University of Hawaii, 2680 Woodlawn Drive, \\
Honolulu, HI 96822, USA}

\author{Curtis N. DeWitt}
\affiliation{University of California, Davis,\\
Phys 539, Davis, \\
CA 95616, USA}

\author{Edward Montiel}
\affiliation{University of California, Davis,\\
Phys 539, Davis, \\
CA 95616, USA}

\author[0000-0002-8594-2122]{Matthew J. Richter}
\affiliation{University of California, Davis,\\
Phys 539, Davis, \\
CA 95616, USA}

\author{Nick Indriolo}
\affiliation{Space Telescope Science Institute,\\
Baltimore, \\
MD 21218, USA}

\author{David A. Neufeld}
\affiliation{Johns Hopkins University,\\
Baltimore, \\
MD 21218, USA}

\author{Yvonne Pendleton}
\affiliation{NASA Ames Research Center,\\
Moffett Field, \\
CA 94035, USA}

\author{Jean Chiar}
\affiliation{Diablo Valley College,\\
321 Golf Club Rd, Pleasant Hill,\\
CA 94523, USA}

\author[0000-0001-6669-0217]{Ryan Dungee}
\affiliation{Institute for Astronomy \\
University of Hawaii, 2680 Woodlawn Drive, \\
Honolulu, HI 96822, USA}

\author{Alexander G. G. M. Tielens}
\affiliation{Leiden University \\
Niels Bohrweg 2, 2333 CA Leiden, \\
The Netherlands}
\nocollaboration

\begin{abstract}

We have performed a 5-8 $\mu$m spectral line survey of the hot molecular core associated with the massive protostar AFGL 2591, using the Echelon-Cross-Echelle Spectrograph (EXES) on the Stratospheric Observatory for Infrared Astronomy (SOFIA). We have supplemented these data with a ground based study in the atmospheric M band around 4.5 $\mu$m using the iSHELL instrument on the Infrared Telescope Facility (IRTF), and the full N band window from 8-13 $\mu$m using the Texas Echelon Cross Echelle Spectrograph (TEXES) on the IRTF. 

Here we present the first detection of ro-vibrational transitions of CS in this source. The absorption lines are centred on average around -10 kms$^{-1}$ and the line widths of CS compare well with the hot component of $^{13}$CO (around 10 kms$^{-1}$). Temperatures for CS, hot $^{13}$CO and $^{12}$CO v=1-2 agree well and are around 700 K. We derive a CS abundance of 8$\times$10$^{-3}$ and 2$\times$10$^{-6}$ with respect to CO and H$_2$ respectively. This enhanced CS abundance with respect to the surrounding cloud (1$\times$10$^{-8}$) may reflect sublimation of H$_2$S ice followed by gas-phase reactions to form CS.

Transitions are in LTE and we derive a density of $>$10$^7$ cm$^{-3}$, which corresponds to an absorbing region of $<$0.04$''$. EXES observations of CS are likely to probe deeply into the hot core, to the base of the outflow. Submillimeter and infrared observations trace different components of the hot core as revealed by the difference in systemic velocities, line widths and temperatures, as well as the CS abundance.

\end{abstract}


\section{Introduction} \label{sec:intro}

Hot cores are objects that are formed during the embedded phase of high mass star formation. They are compact ($\leq$0.1 pc) regions of dense molecular gas with temperatures and densities $\geq$100 K and $\geq$10$^7$ cm$^{-3}$ respectively \citep{Kurtz2000}. They are thought to be an intermediary stage in the formation of massive stars, which starts with the formation of an infrared dark cloud, which then collapses. Once a protostar is formed, it will heat its surroundings, evaporating ice mantles accrued during the preceding, cold, dark cloud phase. The resulting rich organic inventory gives rise to a dense forest of spectral lines in the sub-millimeter (sub-mm) of species such as methanol, dimethyl ether, methyl formate and many others \citep{Blake1987, Plambeck1987, Cesaroni2005}. Regions of warm dense gas with a similarly rich organic inventory have also been observed around low mass stars, where they been been dubbed hot corinos \citep{Ceccarelli2008}.
 
The physics and chemistry of the hot core phase of massive star formation has been well studied in the sub-mm regime by the analysis of pure rotational line emission at high spectral resolution \citep{Kurtz2000, vanderTak2003(1), Beuther2007}. However this is not the case for the mid-infrared (MIR) spectral range, which contains many strong ro-vibrational transitions of important molecules, such as CH$_4$ and C$_2$H$_2$, which do not have transitions in the sub-mm \citep{Lacy1991}. 

High resolution infrared (IR) studies have been carried out before by \citet{Mitchell1990} to analyse CO absorption in a sample of hot cores, including AFGL 2591. Also, \citet{Knez2009} studied the hot core NGC 7538 IRS 1 with TEXES, the Texas Echelon Cross Echelle Spectrograph \citep{Lacy2002}. They detected CS in absorption, however only 6 lines were observed, as well as HNCO and CH$_3$ for the first time in the IR. \citet{Indriolo2015} studied H$_2$O absorption towards AFGL 2591 with the Echelon-Cross-Echelle Spectrograph (EXES; \citeauthor{Richter2010} \citeyear{Richter2010}).

The results presented in this paper are part of the first unbiased spectral survey of the 4.5-13 ${\mu}$m region of hot cores, at high spectral resolution (R=50,000). The high velocity resolution (6 kms$^{-1}$) allows level specific column densities to be determined, and dynamics studied, for the molecules that are detected. This kind of method has come into reach with instruments such as TEXES and iSHELL \citep{Rayner2016} on the Infrared Telescope Facility (IRTF), and EXES onboard the Stratospheric Observatory for Infrared Astronomy (SOFIA; Young et al. 2012).

In this paper we present the first detection of ro-vibrational transitions of CS in AFGL 2591 at MIR wavelengths. CS is well studied in hot cores at sub-mm wavelengths \citep{Li2015, Tercero2010, vanderTak2003}. Sulphur-bearing molecules are very sensitive to the physical conditions in hot cores and therefore provide good tracers for hot core evolution \citep{Hatchell1998}. Sulphur is also known to be heavily depleted in dense regions \citep{Tieftrunk1994} and large discrepancies exist between abundances derived from IR and sub-mm observations \citep{vanderTak2003, Keane2001}. CS therefore provides a good candidate for investigating these issues.

\section{Observations and Data Reduction} \label{sec:obs}

 AFGL 2591 (RA=20:29:24.80, Dec=+40:11:19.0; J2000) was observed with the EXES spectrometer on the SOFIA flying observatory as part of program \texttt{05\_0041} using the high resolution echelon with the low resolution cross disperser (in order 4). Two settings were observed to cover the wavelength range (7.57-7.82 $\mu$m) containing the absorption lines of CS reported here. The slit length and width were 3.2$''$ and 2.89$''$, respectively. The spectral resolving power was R$=$55,000, and the sampling 16 points per resolution element. The observations were done on flights with mission identifications \texttt{2017-03-17\_EX\_F388} and \texttt{2017-03-23\_EX\_F391} at UT 09:51-10:51, latitudes of ~45 degrees, longitudes of -104 degrees, altitudes of 42,000-44,000 feet, and airmasses of 1.70-1.81. In order to remove background emission, the telescope was nodded between the target position and a position 11$''$ to the West and 10$''$ North. 
 
The EXES data were reduced with the SOFIA Redux pipeline \citep{Clarke2015}, which has incorporated routines originally developed for TEXES \citep{Lacy2002}. The science frames were de-spiked and sequential nod positions subtracted, to remove telluric emission lines and telescope/system thermal emission. An internal blackbody source was observed for flat fielding and flux calibration, and the data were divided by the flat field and then rectified, aligning the spatial and spectral dimensions. The wavenumber calibration was carried out from first examining the sky emission spectrum. This allowed for an easier match to the ATRAN wavenumbers in order to calibrate the dispersion and wavenumber zeropoint. There are two wavenumber solutions as the two settings were taken on different days, and they are accurate to 0.42 km s$^{-1}$ and 0.89 kms$^{-1}$. The spectral orders were not flat as a result of instrument fringes. To correct for this the spectrum was divided over a 201 pixel median-smoothed version of itself. Telluric lines were corrected for using an ATRAN model which was not scaled to the depth of the observed telluric lines. The final signal-to-noise values are 130 and 157 at the native sampling.

AFGL 2591 was observed with iSHELL at the IRTF telescope on Maunakea on UT 2017-07-05 from 13:40 to 15:22 at an airmass range of 1.16-1.50, during good weather conditions as part of program 2017A985. The 0.375$''$ slit provided a spectral resolving power of 80,000. The combination of the M1 and M2 configurations provide full coverage from 4.51-5.24 $\mu$m. The target was nodded along the 15$''$ long slit to be able to subtract background emission from the sky and hardware. The total on-target time was 20 minutes for M1 and 31 minutes for M2. iSHELL's internal lamp was used to obtain flat field images. The spectra were reduced with the Spextool package version 5.0.1 \citep{Cushing2004}. Correction for telluric absorption lines was not done with a standard star but with the program xtellcor\_model \footnote{http://irtfweb.ifa.hawaii.edu/research/dr\_resources/} newly developed at the IRTF and makes use of the atmospheric models calculated by the Planetary Spectrum Generator \citep{Villanueva2018}. The Doppler shift of AFGL 2591 at the time of the observations was -35 kms$^{-1}$, and thus telluric and target CO lines are well separated. The blaze shape of the echelle orders were corrected using the flat fields. The final signal-to-noise is ~200 at the native sampling of 2 pixels per resolution element.

\section{Results} \label{sec:results}

\subsection{CS and CO} \label{sec:cite}

CS is a linear molecule and, like CO, it has a simple ro-vibrational spectrum with approximately equidistant lines separated by 2.0 cm$^{-1}$. We have detected 18 CS absorption lines with excitation energies ranging from 14 to 1317 K. For $^{13}$CO, 16 lines were detected with an energy range of 5 to 719 K and 8 lines of the $^{12}$CO v=1-2 band are detected spanning an energy range of 3200 to 4234 K. Missing transitions coincide with strong telluric absorption lines or are blended with other hot core lines. 

\begin{figure}[ht!]
\includegraphics[width=75mm,scale=0.75]{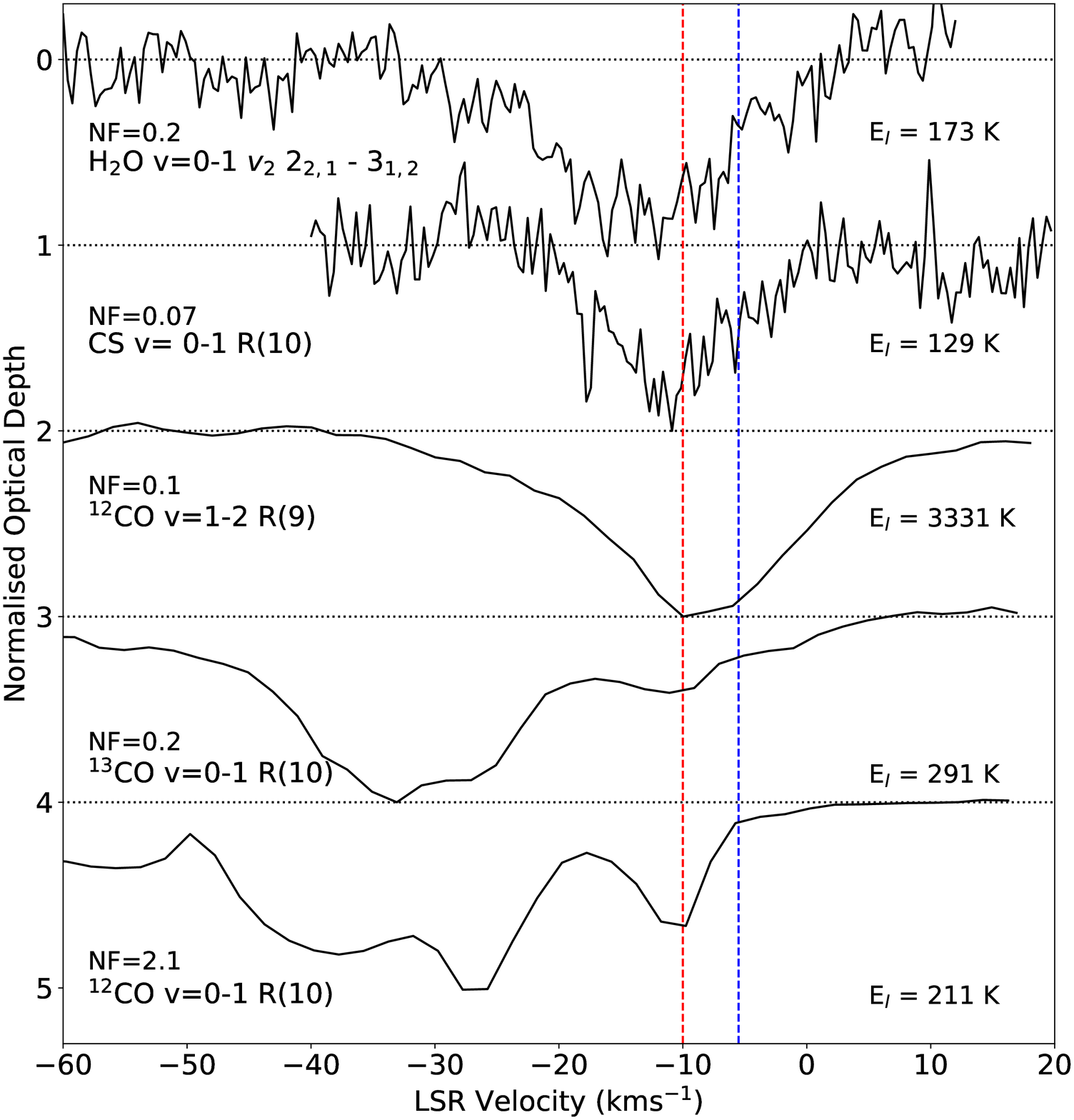}
\includegraphics[width=75mm,scale=0.75]{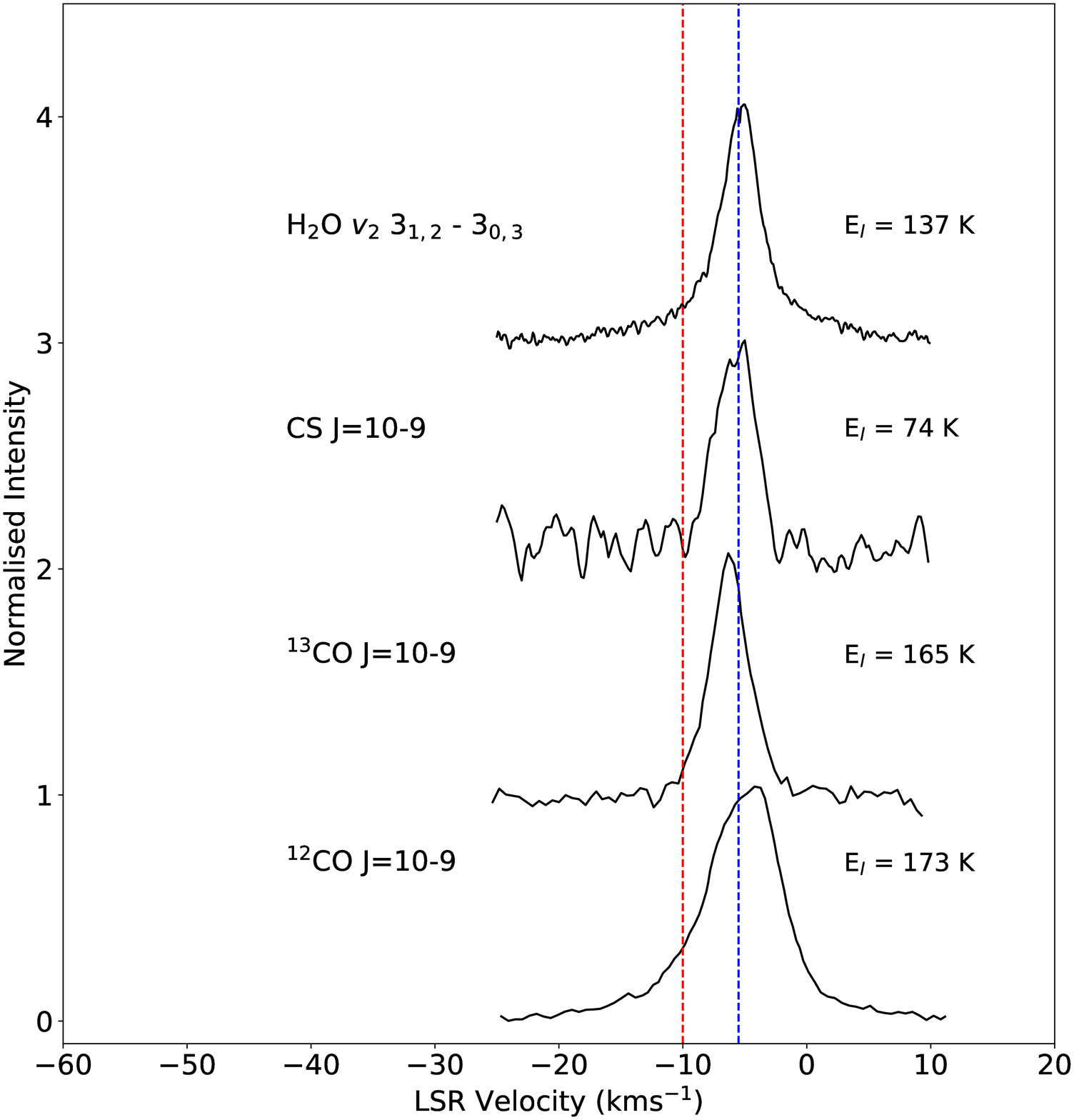}
\caption{Comparison between IR and sub-mm line profiles. The IR lines are shown as optical depth profiles. The sub-mm profiles for CS, $^{13}$CO and $^{12}$CO are taken from \citet{Kazmierczak2014} and the sub-mm H$_2$O line is taken from \citet{Choi2015}. The IR H$_2$O line is from our EXES data. The dashed red and blue lines indicate -10 kms$^{-1}$ and -5.5 kms$^{-1}$ respectively. Lines are normalised and lower energy levels given. For the IR lines, normalisation factors (NF) are given and horizontal baselines are shown.
}
\label{lineeg}
\end{figure}

Figure \ref{lineeg} compares molecular line profiles between the sub-mm and IR in AFGL 2591. The CS lines reveal the presence of a single component at approximately -10 kms$^{-1}$. We have fitted those observed lines with a Gaussian line profile leaving the width, peak position and integrated strength as free parameters.

The $^{12}$CO v=0-1 ro-vibrational lines are saturated up until J=9, and thus we have elected to focus here on the optically thin $^{13}$CO and $^{12}$CO v=1-2 transitions. All components in the $^{12}$CO v=0-1 line blend into one saturated line except for a broad high velocity component indicative of a high velocity outflow.$^{12}$CO v=1-2 line profiles also show a single component at -10 kms$^{-1}$. The line parameters derived from the Gaussian fits are summarised in Table 1. The line widths have been de-convolved with the instrumental resolution.

\begin{figure}[h!]
\centering
\includegraphics[width=75mm,scale=1.5]{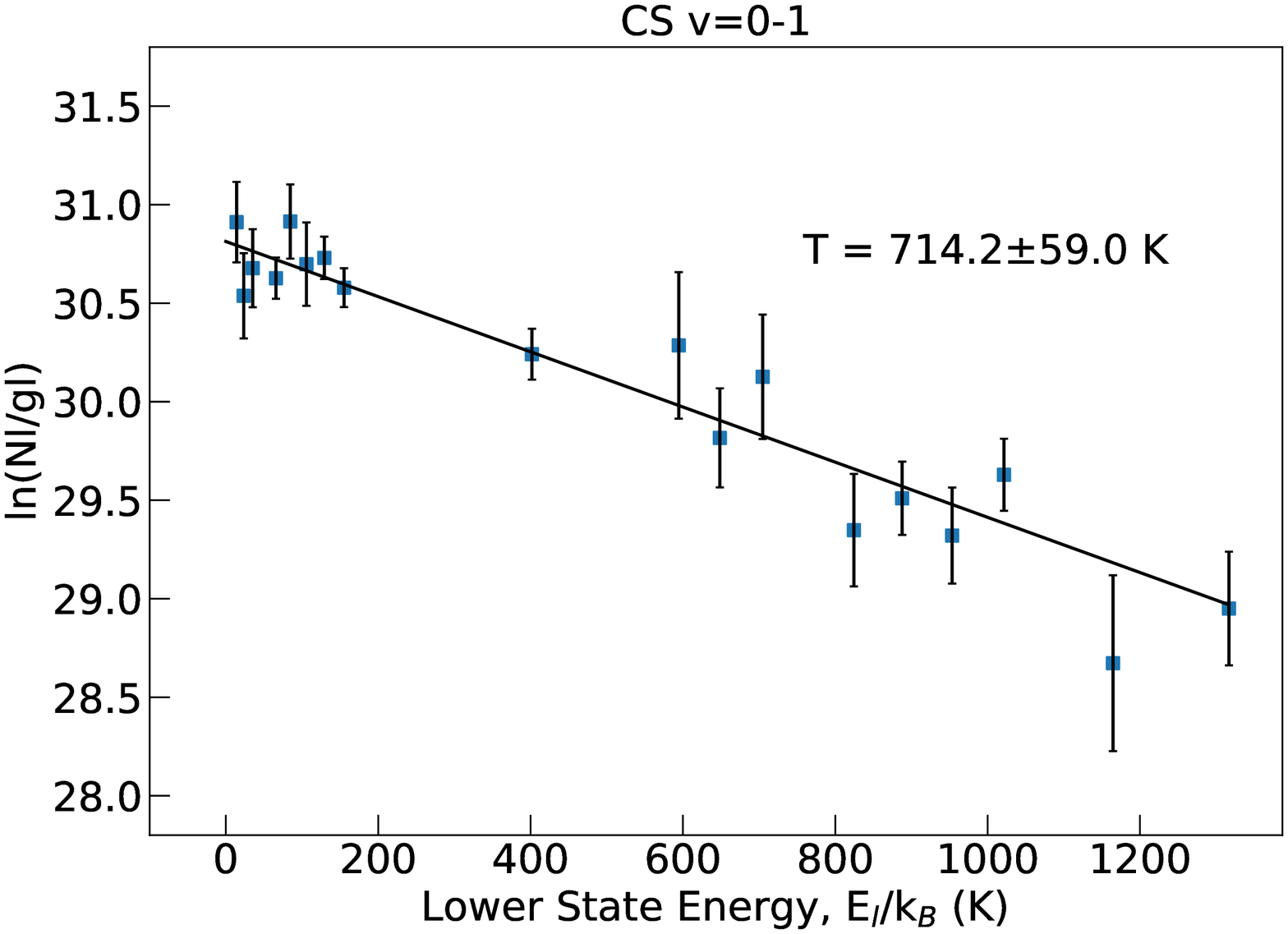}
\includegraphics[width=75mm,scale=1.5]{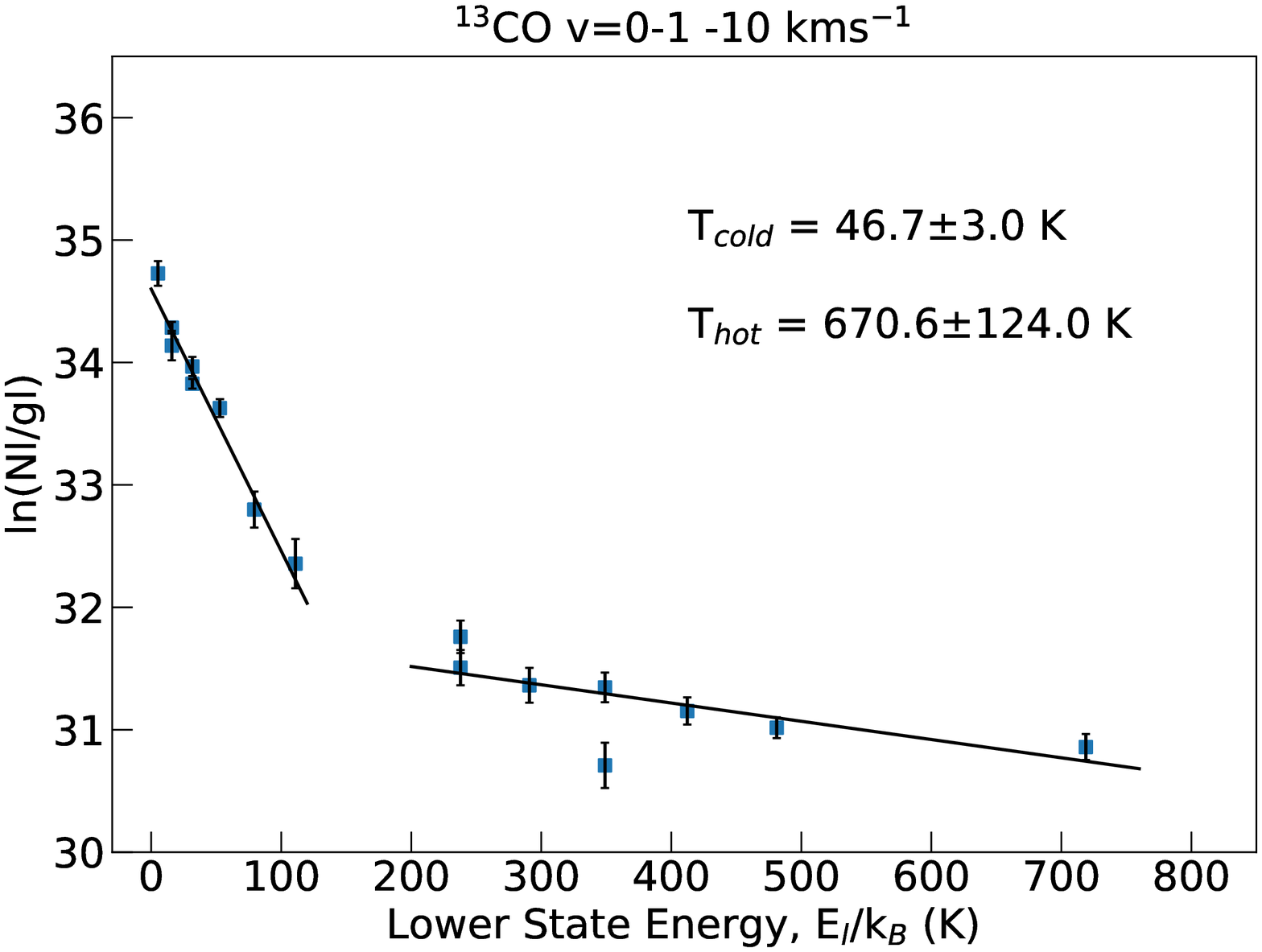}
\includegraphics[width=75mm,scale=1.5]{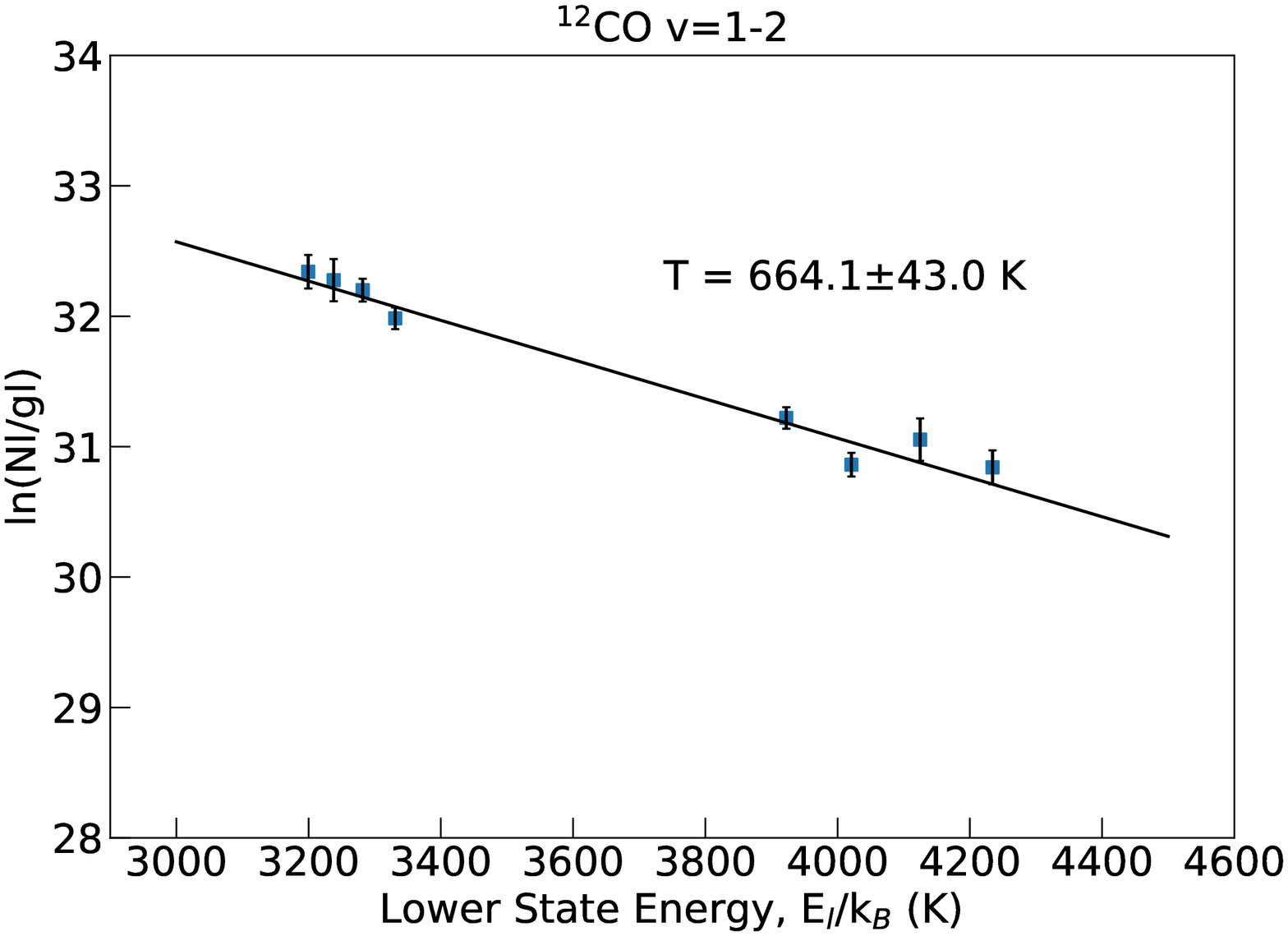}
\includegraphics[width=75mm,scale=1.5]{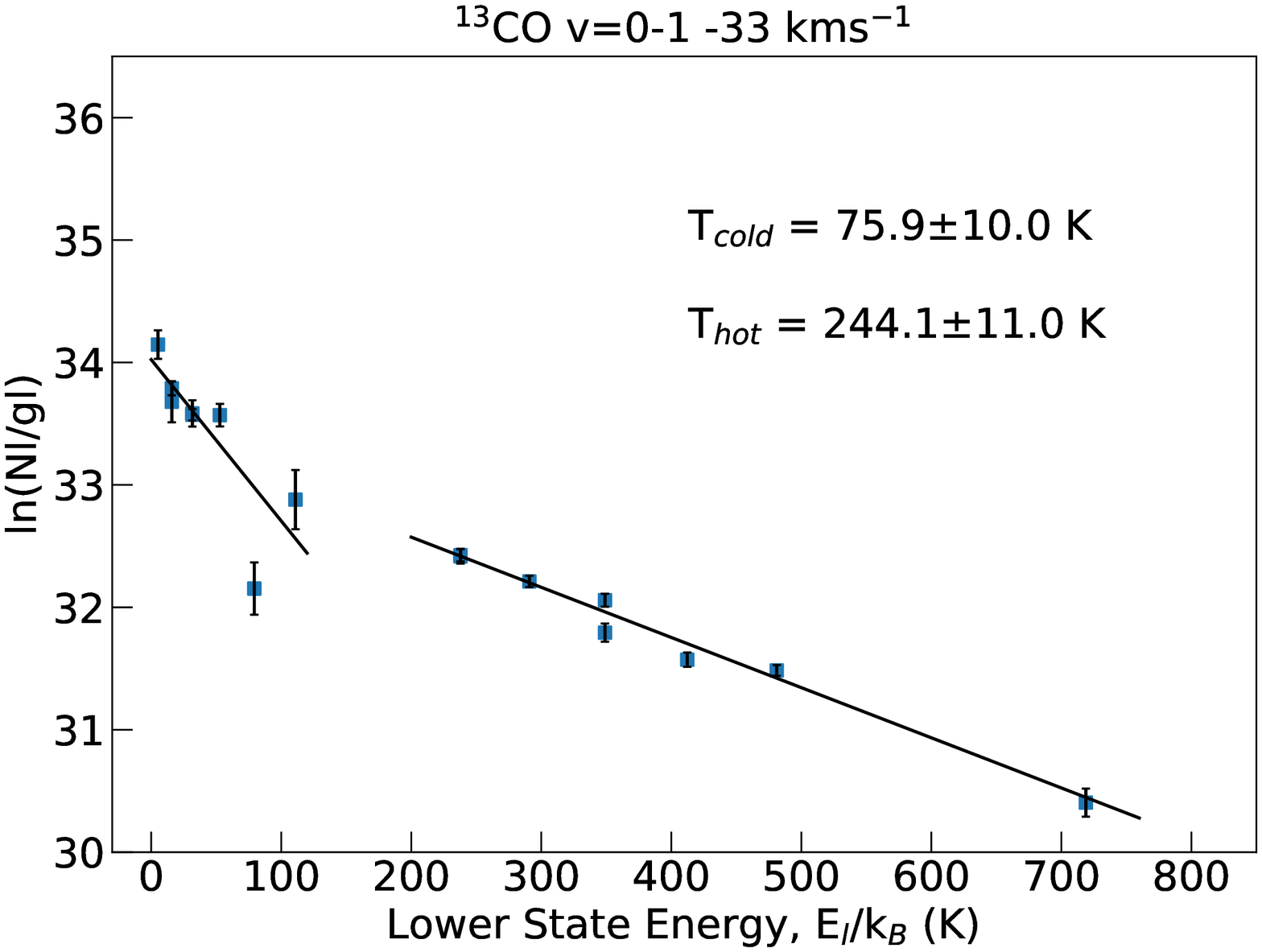}
\caption{CS and CO rotation diagrams. Temperatures for each species are shown, with two temperature components for $^{13}$CO.}
\label{RD}
\end{figure}

The $^{13}$CO lines show complex profiles with multiple velocity components (Figure \ref{lineeg}), one of which coincides in velocity with the CS lines and $^{12}$CO v=1-2 transitions. Two temperature components (hot and cold; see Figure \ref{RD}) are observed at this velocity. For the hot component (high J), we adopted the CS line width but the line width for the cold component (low J) had to be reduced (to 4.1 kms$^{-1}$). Absorption by the narrower cold component overwhelms the contribution by the warm component at the lower J-levels.

\begin{deluxetable*}{cccccCcccc}[h!]
\tablecaption{CS and CO line parameters. ${\Delta}v$ is the line FWHM, ${\tau}_0$ is the optical depth at line centre and N$_l$ is the column density in the lower energy level of the transition, with energy E$_l$. A$_{ij}$ is the Einstein A coefficient for the transition and g$_l$ is the statical weight. 
}
\tablecolumns{11}
\tabletypesize{\scriptsize}
\tablehead{
\colhead{Species} & \colhead{Transition} &
\colhead{Wavelength (${\mu}m$)} & \colhead{E$_l$ (K) } & \colhead{g$_l$} & \colhead{A$_{ij}$ (s$^{-1}$)}  &\colhead{$v_{lsr}$ (kms$^{-1}$)} & \colhead{${\Delta}v$ (kms$^{-1}$)} & \colhead{${\tau}_0$} & \colhead{N$_l$ ($\times$10$^{14}$ cm$^{-2}$)}
}
\startdata
CS v=0-1 & R(3) & 7.8212 & 14.1 & 7 & 7.2 & -11.9 $\pm0.5$ & 8.3 $\pm1.5$ & 0.031 $\pm 0.004$ & 1.9 $\pm0.4$ \\
& R(4) * & 7.8116 & 23.5 & 9 & 7.4 & -9.4 $\pm0.5$ & 6.1 $\pm1.3$ & 0.032  $\pm0.004$ & 1.6 $\pm0.4$ \\
& R(5) & 7.8020 & 35.3 & 11 & 7.6 & -10.1 $\pm0.5$ & 7.7 $\pm1.3$ & 0.039  $\pm 0.004$ & 2.3 $\pm0.5$\\
& R(7) * & 7.7833 & 65.8 & 15 & 7.8 & -10.4 $\pm 0.2$ & 6.4 $\pm 0.6$ & 0.056  $\pm 0.003$ & 3.0 $\pm0.3$ \\
& P(8) & 7.9443 & 84.6 & 17 & 8.1 & -10.9 $\pm0.5$ & 9.6 $\pm1.5$ & 0.055 $\pm 0.006$ & 4.5 $\pm0.9$\\ 
& R(9) & 7.7649 & 105.8 & 19 & 8.0 & -9.5 $\pm0.6$ & 11.2 $\pm1.8$ & 0.051 $\pm0.005$ & 4.1 $\pm0.5$\\
& R(10) & 7.7559 & 129.2 & 21 & 8.0 & -11.1 $\pm0.3$ & 10.4 $\pm0.9$ & 0.061  $\pm 0.004$& 4.7 $\pm0.5$ \\
& R(11) & 7.7469 & 155.1 & 23 & 8.1 & -12.1 $\pm0.3$ & 9.9 $\pm0.8$ & 0.060  $\pm 0.002$ & 4.4 $\pm0.4$\\
& R(18) & 7.6868 & 401.8 & 37 & 8.5 & -12.0 $\pm0.3$ & 11.0 $\pm0.8$ & 0.063  $\pm 0.003$ & 5.0 $\pm0.6$ \\
& R(22) * & 7.6545 & 594.5 & 45 & 8.7 & -12.7 $\pm0.4$ & 10.7 $\pm1.2$ & 0.080  $\pm 0.004 $& 6.1 $\pm2.4$\\
& R(23) & 7.6467 & 648.5 & 47 & 8.8 & -10.3 $\pm0.9$ & 8.4 $\pm3.8$ & 0.016  $\pm 0.003 $& 4.2 $\pm 1.1$\\
& R(24) * & 7.6389 & 704.7 & 49 & 8.8 & -5.2  $\pm0.7$ & 15.0 $\pm1.8$ & 0.060  $\pm 0.003$& 5.9 $\pm1.9$\\
& R(26) & 7.6237 & 824.4 & 53 & 8.9 & -8.4 $\pm0.6$ & 7.9 $\pm2.0$ & 0.048 $\pm0.007$ & 3.0 $\pm0.7$\\
& R(27) * & 7.6162 & 887.8 & 55 & 9.0 & -11.2 $\pm0.6$ & 4.9 $\pm1.2$ & 0.076 $\pm0.006$ & 3.6 $\pm0.8$\\
&  R(28) & 7.6088 & 953.4 & 57 & 9.0 & -10.0 $\pm0.4$ & 6.8 $\pm1.0$ & 0.055  $\pm 0.008 $& 3.1 $\pm0.7$\\
& R(29) & 7.6015 & 1021.5 & 59 & 9.1 & -10.9 $\pm0.4$ & 6.7 $\pm1.2$ & 0.078 $\pm0.008$ & 4.4 $\pm0.8$\\
& R(31) * & 7.5871 & 1164.5 & 63 & 9.2 & -8.0 $\pm0.9$ & 11.4 $\pm4.1$ & 0.024 $\pm0.005$ & 1.8 $\pm0.8$\\
&  R(33) & 7.5731 & 1316.8 & 67 & 9.2 & -9.0 $\pm0.6$ & 5.2 $\pm1.5$ & 0.051  $\pm 0.009 $& 2.5 $\pm0.7$ \\ 
\hline
$^{13}$CO v=0-1 & P(1) & 4.7792 & 5.2 & 6 & 32.4 & -9.4 $\pm0.2$ & 4.1 & 0.50 $\pm0.04$ & 72.4 $\pm7.3$\\
&P(2) & 4.7877 & 15.8 & 10 & 21.5 & -9.3 $\pm0.1$ & 4.1 & 0.57 $\pm0.02$ & 77.5 $\pm3.6$ \\
& R(2) & 4.7463 & 15.8 & 10 & 14.2 & -9.7 $\pm0.3$ & 4.1 & 0.68 $\pm0.07$ & 67.0 $\pm8.0$ \\
&R(3) & 4.7383 & 31.7 & 14 & 14.8 & -9.3 $\pm0.1$ & 4.1 & 0.70 $\pm0.02$ & 68.6 $\pm2.7$ \\
& P(3) & 4.7963 & 31.7 & 14 & 19.3 & -9.3 $\pm0.2$ & 4.1 & 0.61 $\pm0.04$ & 79.0 $\pm6.2$ \\
& P(4) & 4.8050 & 52.8 & 18 & 18.2 & -9.3 $\pm0.2$ & 4.1 & 0.57 $\pm0.03$ & 72.4 $\pm5.2$ \\
&R(5) & 4.7227 & 79.3 & 22 & 15.5 & -8.3 $\pm0.2$ & 4.1 & 0.44 $\pm0.06$ & 38.6 $\pm5.7$ \\
& R(6) & 4.7150 & 111.0 & 26 & 15.8 & -9.1 $\pm0.3$ & 4.1 & 0.31 $\pm0.06$ & 29.3 $\pm5.9$ \\
& R(9) & 4.6927 & 237.9 & 38 & 16.4 & -12.1 $\pm0.7$ & 11.2 & 0.14 $\pm0.01$ & 23.6 $\pm3.1$ \\
& P(9) & 4.8501 & 237.9 & 38 & 16.4 & -13.2 $\pm0.8$ & 11.2 & 0.10 $\pm0.01$ & 18.3 $\pm2.6$ \\
& R(10) & 4.6853 & 290.8 & 42 & 16.5 & -10.5 $\pm0.6$ & 11.2 & 0.10 $\pm0.01$ & 17.5 $\pm2.5$ \\
& R(11) & 4.6782 & 349.0 & 46 & 16.7 & -12.0 $\pm0.5$ & 11.2 & 0.11 $\pm0.01$ & 18.9 $\pm2.3$ \\
&P(11) & 4.8689 & 349.0 & 46 & 16.1 & -13.0 $\pm0.6$ & 11.2 &  0.05 $\pm0.01$ & 10.0 $\pm1.8$ \\
& R(12) & 4.6711 & 412.3 & 50 & 16.8 & -12.0 $\pm0.6$ & 11.2 & 0.09 $\pm0.01$ & 16.9 $\pm1.9$ \\
& R(13) & 4.6641 & 481.1 & 54 & 16.9 & -12.0 $\pm0.4$ & 11.2 & 0.09 $\pm0.01$ & 15.9 $\pm1.4$ \\
& R(16) & 4.6437 & 718.8 & 66 & 17.3 & -10.8 $\pm0.5$ & 11.2 & 0.10 $\pm0.01$ & 16.6 $\pm 1.8$ \\
\hline
$^{12}$CO v=1-2 & R(6) & 4.6675 & 3199.5 & 13 & 33.2 & -9.6 $\pm0.6$ & 17.9 $\pm1.4$ & 0.102 $\pm0.006$ & 13.1 $\pm1.6$ \\ 
 & R(7) & 4.6598 & 3237.9 & 15 & 33.7 & -8.2 $\pm0.7$ & 19.9 $\pm1.9$ & 0.108 $\pm0.008$ & 15.6 $\pm2.6$\\
 & R(8) & 4.6522 & 3281.8 & 17 & 34.1 & -9.1 $\pm0.4$ & 18.3 $\pm0.9$ & 0.124 $\pm0.006$ & 16.4 $\pm1.4$ \\
 & R(9) & 4.6448 & 3331.0 & 19 & 34.4 & -8.6 $\pm0.4$ & 17.9 $\pm0.9$ & 0.113 $\pm0.005$ & 14.8 $ \pm1.2$\\
 & R(17) & 4.5887 & 3922.5 & 35 & 36.6 & -8.5 $\pm0.3$ & 16.6 $\pm0.9$ & 0.104 $\pm0.005$ & 12.7 $\pm1.1$\\
 & P(18) & 4.8948 & 4020.9 & 37 & 31.7 & -7.1 $\pm0.3$ & 13.9 $\pm0.9$ & 0.083 $\pm0.004 $& 9.4 $\pm0.9$\\
 &  P(19) & 4.9054 & 4124.8 & 39 & 31.4 & -9.9 $\pm0.8$ & 14.6 $\pm3.8$ & 0.010 $\pm0.009$ & 12.0 $\pm1.9$\\
 &  R(20) & 4.5694 & 4234.4 & 41 & 37.2 & -6.1 $\pm0.5$ & 16.0 $\pm1.2$ & 0.088 $\pm0.005$ & 10.3 $\pm1.5$\\
\enddata
\label{CSandCO}
\tablenotetext{}{Note: *These lines suffer from systematic error such as poor baselines therefore there is a larger uncertainty in the continuum placement. \\ Line data were taken from the HITRAN database \citep{Gordon2017}.}
\end{deluxetable*}

The detected, unblended lines of CS and CO are presented in Figure \ref{linesall} in the Appendix. There is a spread in centroid velocity for a given species. The line widths for CS and hot $^{13}$CO are in agreement for equivalent energy level.

The rotation diagrams of CS, $^{13}$CO and $^{12}$CO v=1-2 are shown in Figure \ref{RD}. In all cases the rotation diagrams are straight lines indicating that local thermodynamic equilibrium (LTE) is a good approximation. We have calculated the vibrational excitation temperature of $^{12}$CO from the v=1-2 transitions and the warm $^{13}$CO column density assuming a $^{12}$CO/$^{13}$CO abundance ratio of 60 \citep{Wilson1994}. The derived vibrational excitation temperature of 625$\pm$56 K agrees well with a similar estimate by \citet{Mitchell1989}. The similar values for the rotational and vibrational temperatures suggest vibrational LTE.

The excitation temperature and velocity for all three species (CS, $^{13}$CO high J and $^{12}$CO v=1-2) are very similar. This supports an origin in the same region. The physical conditions for the detected species are summarised in Table 2. An estimate of the column density of hot $^{12}$CO from the lines of vibrationally excited $^{12}$CO gives 1.5$\pm$0.6$\times$10$^{18}$ cm$^{-2}$. We derive an upper limit on CS in the -33 kms$^{-1}$ velocity component of $<$ 8$\times$10$^{14}$ cm$^{-2}$.

\begin{deluxetable*}{ccccc}[h!]
\tablecaption{Summary of Molecular Species in AFGL 2591\label{tab:pars}}
\tablecolumns{5}
\tabletypesize{\scriptsize}
\tablehead{
\colhead{Species} & \colhead{$v_{lsr}$ (kms$^{-1}$)} & \colhead{T$_{ex}$ (K)} & \colhead{N ($\times$10$^{16}$cm$^{-2}$)} & \colhead{N/N($^{12}$CO)$^1$}
}
\startdata
CS v=0-1 & -10 $\pm$0.5 & 714 $\pm$59 & 1.6 $\pm$0.1 & 8.0$\times$10$^{-3}$\\
$^{12}$CO v=1-2 & -8.4 $\pm$0.5 & 664 $\pm$43 & 1.48 $\pm$0.6 & 7.4$\times$10$^{-3}$ \\
$^{13}$CO v=0-1 & -11.95 $\pm$0.6 & 670 $\pm$124 & 3.4 $\pm$0.4 & 0.02\\
H$_2$O$^2$ & -11 $\pm$0.3 & 640 $\pm$80 & 370 $\pm$80 & 1.85\\
$^{13}$CO v=0-1 & -9.2 $\pm$0.2 & 47 $\pm$3 & 3.8 $\pm$0.2 & 0.02\\
$^{13}$CO v=0-1 & -33 $\pm$0.8 & 76 $\pm$10 & 3.4 $\pm$0.2 & 0.02\\
$^{13}$CO v=0-1 & -33 $\pm$0.5& 244 $\pm$11 & 5.9 $\pm$ 0.4 & 0.03\\
CS v=0-1 & -33 $\pm$0.5 & 244 $\pm$11 & $<$ 0.08 & $<$ 2.2$\times$10$^{-4}$\\
\enddata
\tablenotetext{1}{N($^{12}$CO) = N($^{13}$CO)$\times$60 \citep{Wilson1994}}
\tablenotetext{2}{\citet{Indriolo2015} with EXES}
\label{Summary}
\end{deluxetable*}

\subsection{CS and CO: Infrared vs Submillimeter} \label{sec:cite}

From our observations we derive a different systemic velocity than from sub-mm studies of AFGL 2591 (Figure \ref{lineeg}) in both CS, $^{13}$CO and $^{12}$CO v=1-2 (-10 kms$^{-1}$ compared to -5.5 kms$^{-1}$, \citeauthor{Bally1983}\citeyear{Bally1983}). \citet{vanderTak1999} studied AFGL 2591 at sub-mm, radio and IR wavelengths, at lower resolution, and derive a centroid velocity for $^{13}$CO absorption of -5.5 kms$^{-1}$, in agreement with the systemic velocity of rotational emission lines, including CS \citep{Boonman2001, Benz2007, Kazmierczak2014}. However their IR transitions show a spread in velocity of -3.5 to -12.4 kms$^{-1}$ which they attribute to atmospheric interference, picking only a few lines to derive their value of -5.5 kms$^{-1}$.

The intrinsic line widths that we observed are broader than for sub-mm studies (Figure \ref{lineeg}). As an example, the CS J=10-9 transition has a width of 3.5 kms$^{-1}$ whereas our CS rovibrational transition v=0-1 R(10) has a width of 10.4 kms$^{-1}$. The H$_2$O 3(1,2)-3(0,3) line has a width of 3.1 kms$^{-1}$ in the sub-mm compared to 16.5 kms$^{-1}$ for the ${\nu}_2$=0-1 2(2,1)-3(1,2) IR line. 

We suggest our higher temperature, higher velocity IR observations trace more turbulent gas closer to the central source, probing deep into the base of the outflow, whereas sub-mm observations trace more extended gas at the velocity of the quiescent envelope. 

The CS, $^{13}$CO v=0-1 and $^{12}$CO v=1-2 temperatures are very similar and given the difference in critical densities, suggest gas in LTE. The critical densities for J=7-6 are 2$\times$10$^5$ cm$^{-3}$ \citep{Yang2010} and 3$\times$10$^7$ cm$^{-3}$ \citep{Turner1992} for CO and CS respectively. Therefore we estimate that the density must exceed 10$^{7}$ cm$^{-3}$ in order to maintain LTE. The critical density of $^{12}$CO v=1-2 cannot be used due to the high optical depth of the $^{12}$CO v=0-1 transitions, which may lead to line trapping and tends to increase the vibrational excitation temperature.

Assuming a $^{12}$CO/H$_2$ abundance of 2x10$^{-4}$ \citep{Lacy1994} and constant density of 10$^{7}$ cm$^{-3}$, we derive a physical size for the absorbing region of $<$ 130 AU. At a distance of 3.3 kpc \citep{Rygl2012} this corresponds to $<$0.04$''$.

The velocity structure of the inner region (down to 500 AU) of the hot core in AFGL 2951 has been well resolved and modelled by \citet{Wang2012}. A high velocity gradient is found for SO$_2$ and the velocity field of the blue shifted component is observed to have the highest negative velocities towards the centre on the source. However the highest velocity that is reached is $\sim$ -7 kms$^{-1}$, at the very centre. This further indicates that CS at -10 kms$^{-1}$ would not be observed at sub-mm wavelengths. Combined with the fact that SO$_2$ is a molecular tracer for shocks or outflows \citep{Schilke1997}, \citet{Wang2012} conclude that the SO$_2$ emission originates in an interactive layer between the upper parts of a disk-like structure and the outflow.

\citet{vanderTak1999} conclude that the outflow of AFGL 2591 points towards us along the line of sight. In order to explain CS (7-6) emission, \citet{Bruderer2009 propose that their observations trace dense walls of the outflow, at a velocity of around -5.5 kms$^{-1}$.} Given that the extreme velocities originate in the centre of the source, and that our EXES observations show an average velocity around -10 kms$^{-1}$, we propose that we are looking into the outflow, probing deeply towards the protostar to the base of the outflow. Also the broad line profiles of CS are indicative of a shocked region. The high temperature implies that the gas lies close to the protostar, further supporting the proposed origin as the base of the outflow.

\subsection{Chemistry of CS} \label{sec:cite}

We derive a CS/$^{12}$CO abundance of 8x10$^{-3}$ and CS/H$_2$ abundance of 2$\times$10$^{-6}$, assuming a CO/H$_2$ ratio of 2$\times$10$^{-4}$. This abundance is two orders of magnitude higher than sub-mm observations of the hot core and envelope \citep{vanderTak2003, Jimenez2012}. This, once again, illustrates that the IR observations probe a very different region in this source. Consequently hot CS contains 6$\pm$0.8 \% of the cosmic sulphur budget, assuming an S/H ratio of 1.3$\times$10$^{-5}$. Compared to the Orion Hot Core which has a CS/CO abundance of 4.1$\times$10$^{-4}$ derived from sub-mm observations \citep{Tercero2010}, the CS abundance in the gas traced by the IR observations of the hot core in AFGL 2591 is much higher.

Chemical models based on sub-mm observations \citep{Charnley1997, Doty2002} derive a CS abundance of the order 1$\times$10$^{-8}$ with respect to H$_2$. A two phase, time-dependent gas-grain model \citep{Viti2004} has been used to study the sulphur chemistry in the Orion Hot Core, and predicts the abundance of CS as a function of hot core age \citep{Esplugues2014}. The first and second phase simulate depletion onto, and sublimation from, grain surfaces respectively. For a hot core with a mass of 10 M$_{\odot}$, solar sulphur abundance and a density of 10$^7$ cm$^{-3}$, a CS/CO abundance of around 5$\times$10$^{-3}$ is achieved after 6$\times$10$^4$ yr.
 
The abundances of CS, H$_2$CS and SO$_2$ increase such that these species become the most abundant sulphur-bearing molecules for an evolved hot core \citep{Esplugues2014}. The main chemical pathway that is responsible for the production of CS in this model is CH$_2$ + S $\rightarrow$ CS + H$_2$. Chemical models predict large amounts of atomic sulphur at high temperatures \citep{Doty2002}. This arises due to abstraction of H$_2$S which is efficient at high temperatures. At the same time small hydrocarbons are formed by the breakdown of CO via cosmic ray ionisation. 

Therefore we propose two scenarios to interpret the CS abundance. The first is that AFGL 2591 is a more evolved hot core in which all sulphur is converted to H$_2$S on grain surfaces, and then converted back to S in the gas-phase after ice mantle sublimation. Then, at long timescales, enough CS is produced in the hot inner region of the hot core to explain our observations \citep{Esplugues2014}. These models are not optimised to AFGL 2591 which would likely have an effect on the CS abundance, therefore a more in depth study using chemical models would be necessary to clarify the proposed timescales. We note that H$_2$S ice has not been observed in absorption towards massive protostars \citep{Smith1991} at abundance upper limits a factor of 7 lower than that of CS, putting H$_2$S as a source of the gas phase S into question. Deeper searches for H$_2$S would be very helpful to asses the sulphur budget in interstellar ices.

The alternative scenario is that our observations trace a disk-wind interaction zone very close to the protostar. In this case the cosmic ray ionisation rate would be high, which would favour the breaking of C out of CO, which could lead to the enhancement of CS production. Again, atomic sulphur is produced via abstraction of H$_2$S. \citet{May2000} find that grain sputtering becomes important in shocks around 15 kms$^{-1}$, therefore sulphur could also be released from grain mantles in the presence of shocks.

The conditions and chemical history of AFGL 2591 are clearly favourable for the production of CS. A high CS/CO abundance of 7$\pm0.4$$\times$10$^{-3}$ has also been observed at MIR wavelengths in NGC 7538 IRS 1 \citep{Knez2009} with TEXES. Nevertheless, the enhanced CS abundance is still not enough to appoint this molecule as the main reservoir of sulphur in hot cores. A high abundance of warm SO$_2$ has also been observed with EXES in the hot core Mon R2 IRS 3. (Dungee et al., submitted to ApJL). This also is not observed in the sub-mm suggesting that a large amount of sulphur is visible only at IR wavelengths.

\section{Conclusions}

We present the first detection of ro-vibrational transitions of CS in the hot core of AFGL 2591 with EXES. The CS observations are complemented with high resolution iSHELL CO observations. The CO gas is found to have five velocity components, one of which is consistent with the velocity of CS, -10 kms$^{-1}$. $^{12}$CO v=1-2 is also observed at this velocity. A temperature of 714$\pm59$ K is derived from the rotation diagram of CS, and the observation of CS up to J level of 33, along with a similar excitation temperature for the pure rotational CO lines, imply high densities ($>10^{7}$ cm$^{-3}$). The temperature is consistent with hot $^{13}$CO and $^{12}$CO v=1-2 which have 670$\pm124$ K and 664$\pm43$ K respectively.

The systemic velocity of AFGL 2591 that we derive is 5 kms$^{-1}$ bluer than that derived from sub-mm observations. We propose that this is because we are observing the base of the blue-shifted outflow very close to the central IR source. This is reflected in the high densities and temperatures derived in our observations.

The abundance of CS observed to be 8x10$^{-3}$ and 2$\times$10$^{-6}$ with respect to CO and H$_2$ respectively. This is two orders of magnitude above what is derived from sub-mm observations, 1$\times$10$^{-8}$ with respect to H$_2$. This provides evidence of a large sulphur depository which is detectable more readily at IR wavelengths. IR observations are sensitive to a different region of the hot core than sub-mm observations. IR observations of CS trace gas in the hot core that is much hotter and denser than do sub-mm observations, and that is at a larger systemic velocity. Therefore they probe much deeper into the innermost parts of the hot core, avoiding any contamination by the surrounding envelope.

Chemical models support the derived abundance of CS if AFGL 2591 is an evolved hot core. Alternatively our observations may be tracing the onset of a disk wind at the base of the outflow.

\acknowledgments

Based [in part] on observations made with the NASA/DLR Stratospheric Observatory for Infrared Astronomy (SOFIA). SOFIA is jointly operated by the Universities Space Research Association, Inc. (USRA), under NASA contract NAS2-97001, and the Deutsches SOFIA Institut (DSI) under DLR contract 50 OK 0901 to the University of Stuttgart. A.G.G.M.T thanks the Spinoza premie of the NWO. D.A.N gratefully acknowledges the support of an USRA SOFIA grant, SOF05-0041.

\section{Appendix}

\begin{figure}[b!]
\centering
\includegraphics[width=45mm,scale=1.5]{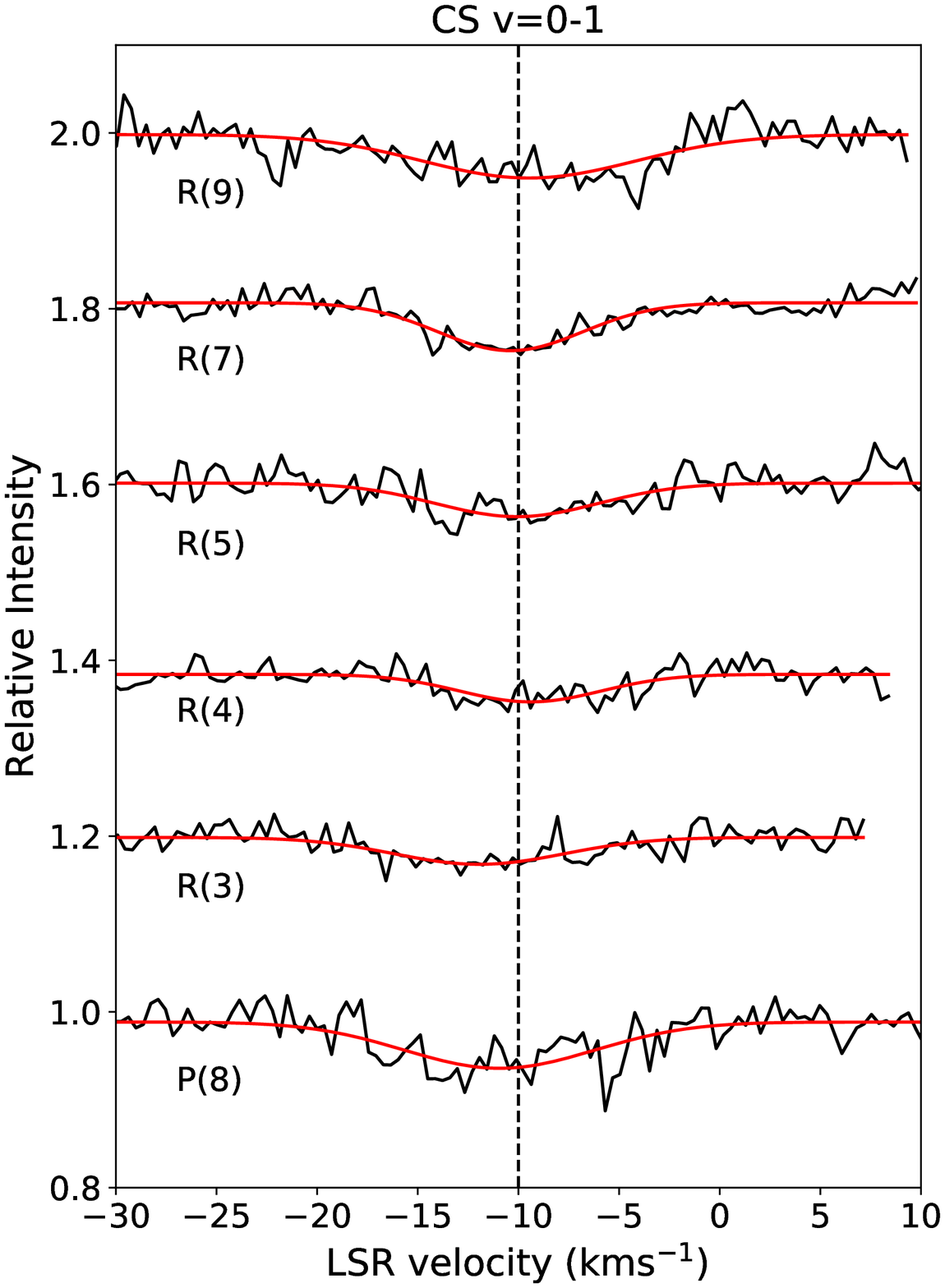}
\includegraphics[width=45mm,scale=1.5]{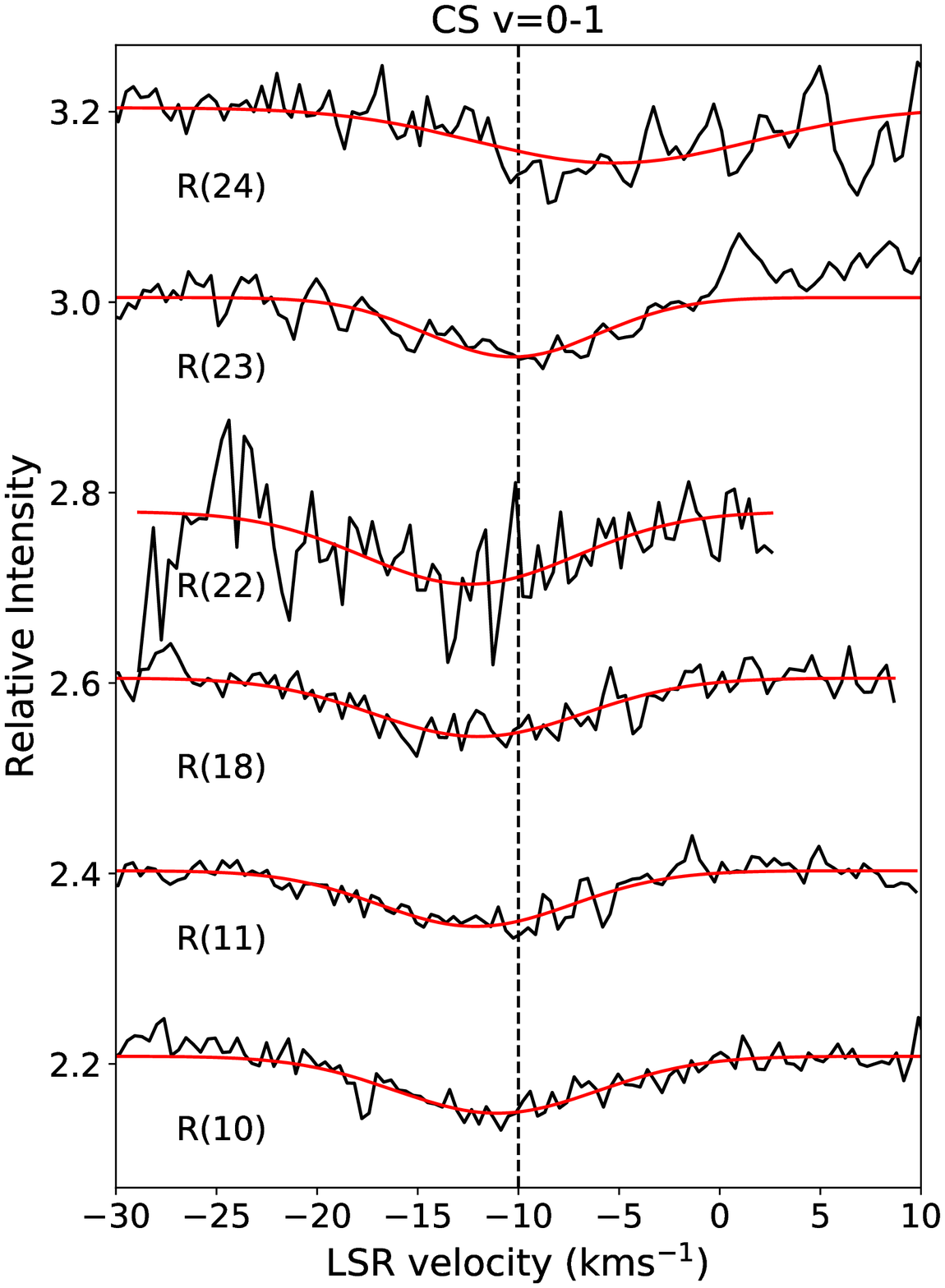}
\includegraphics[width=45mm,scale=1.5]{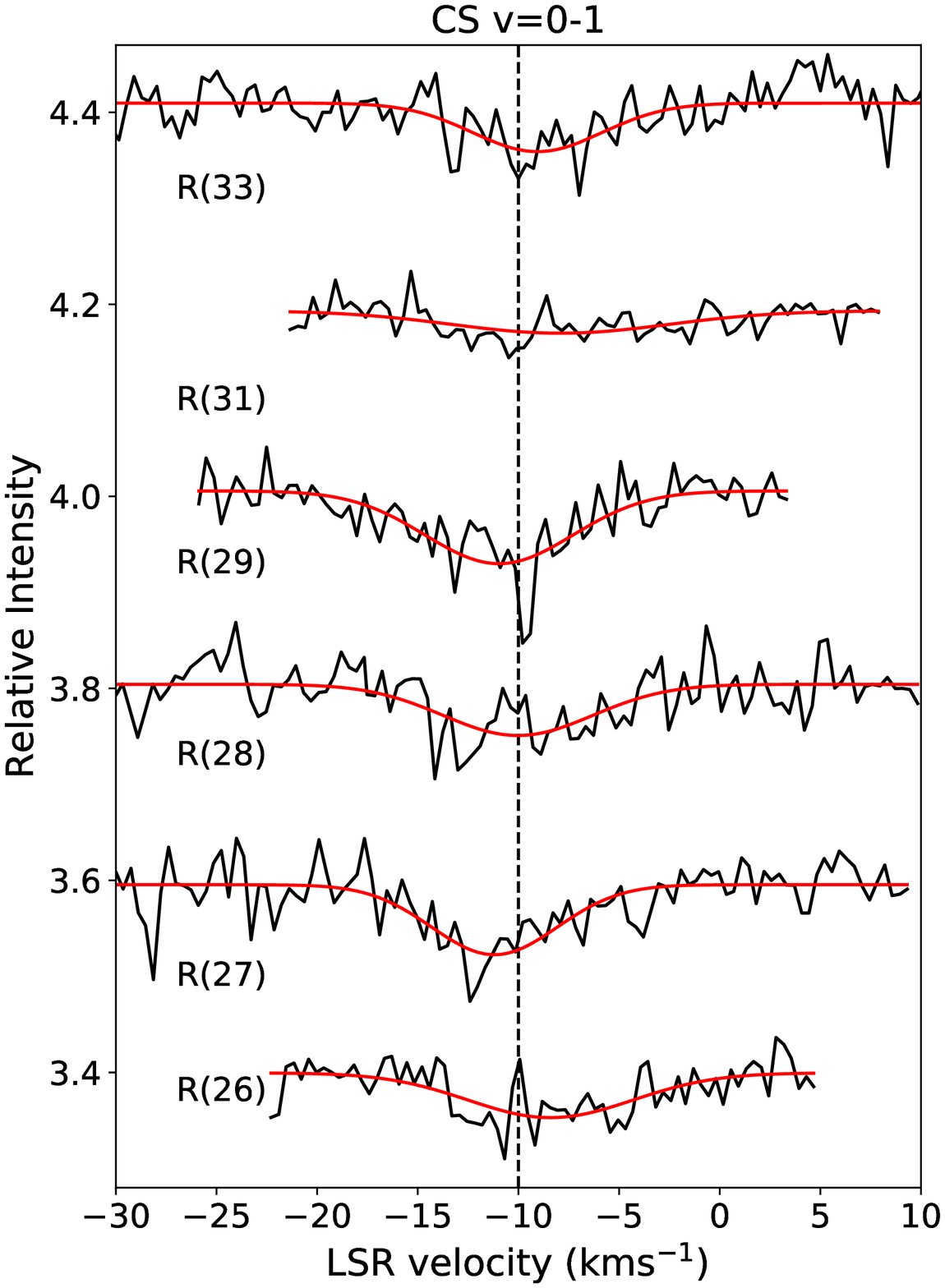}\\
\includegraphics[width=45mm,scale=1.5]{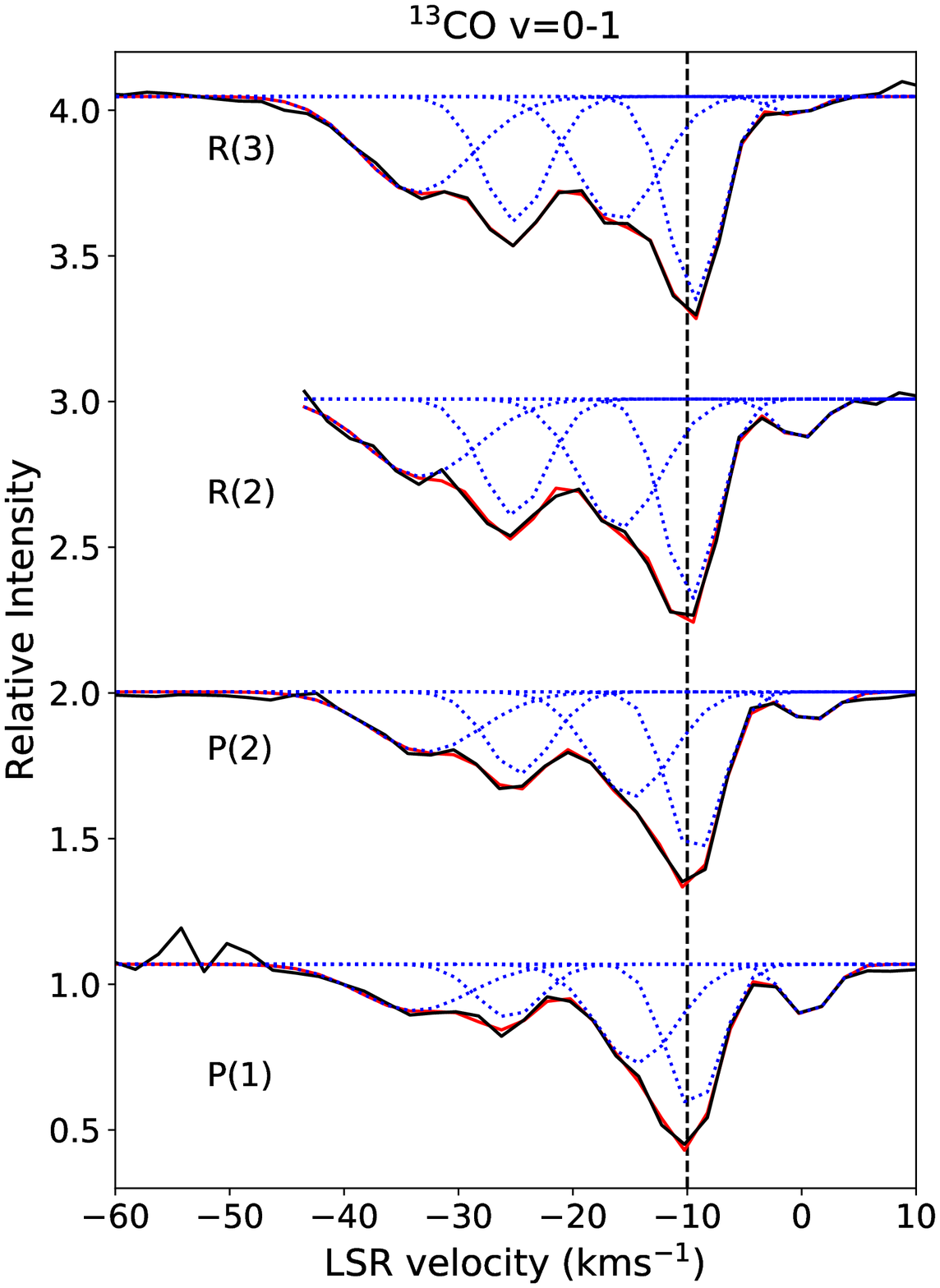}
\includegraphics[width=45mm,scale=1.5]{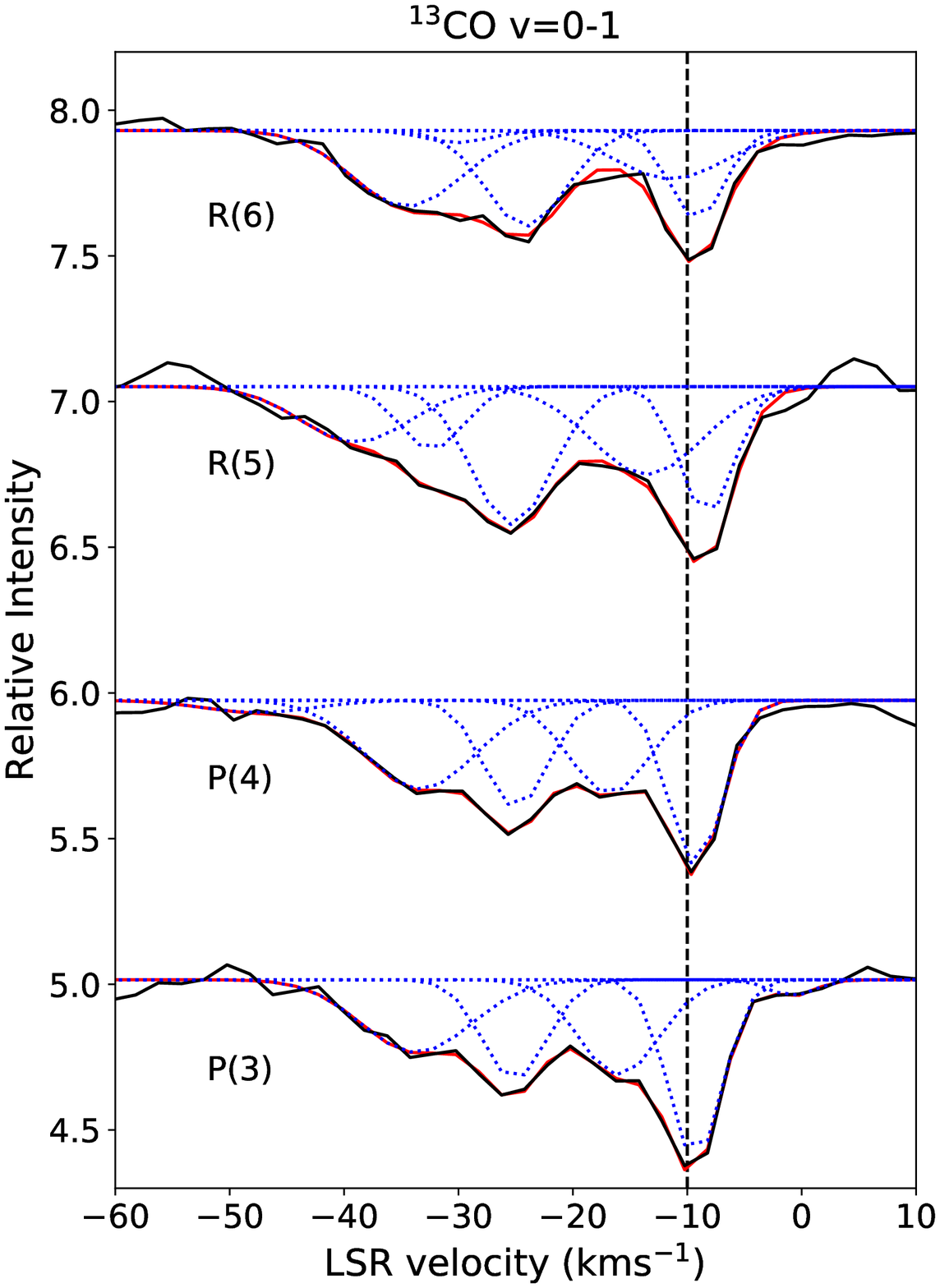}
\includegraphics[width=45mm,scale=1.5]{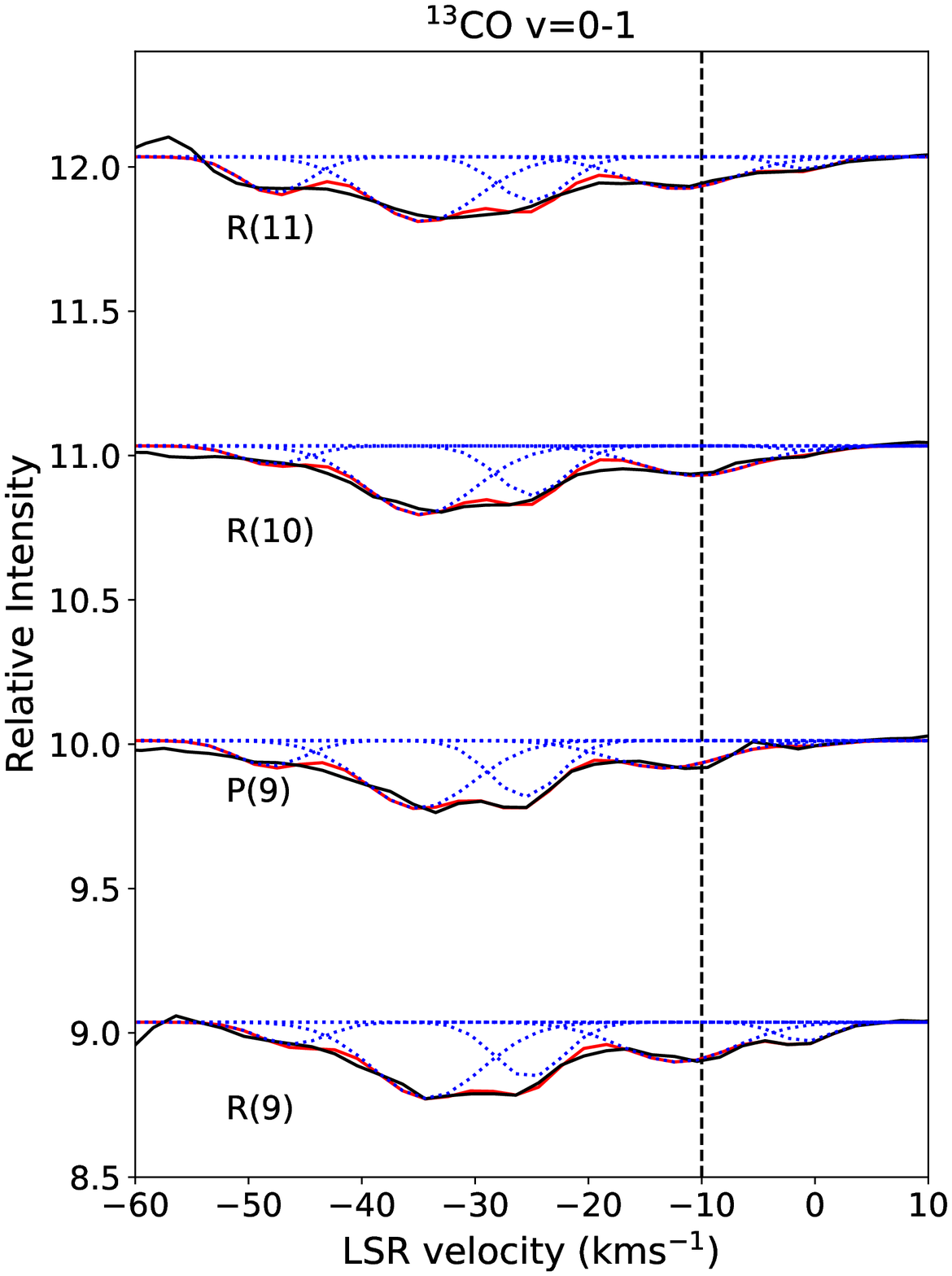}\\
\includegraphics[width=45mm,scale=1.5]{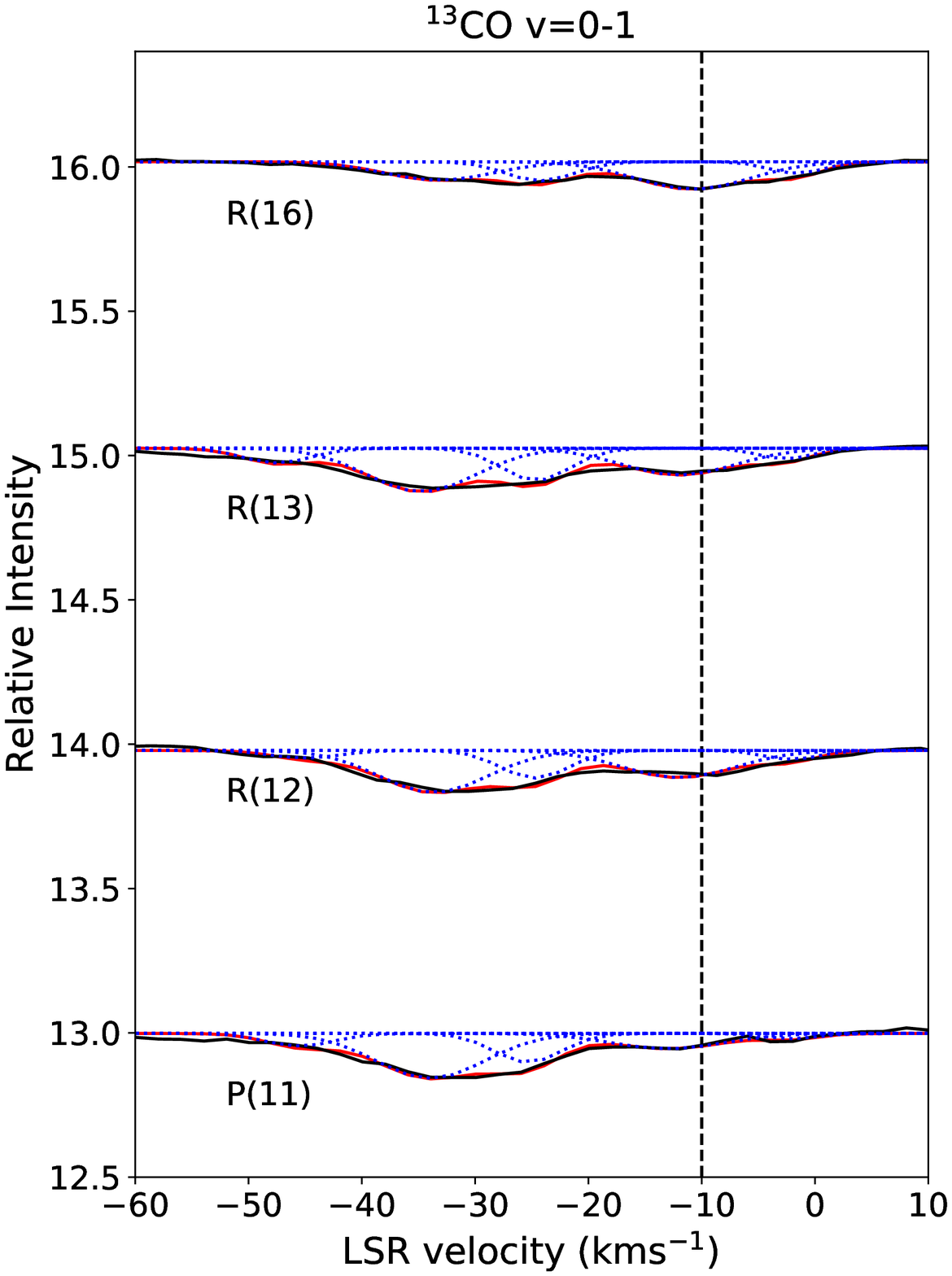}
\includegraphics[width=45mm,scale=1.5]{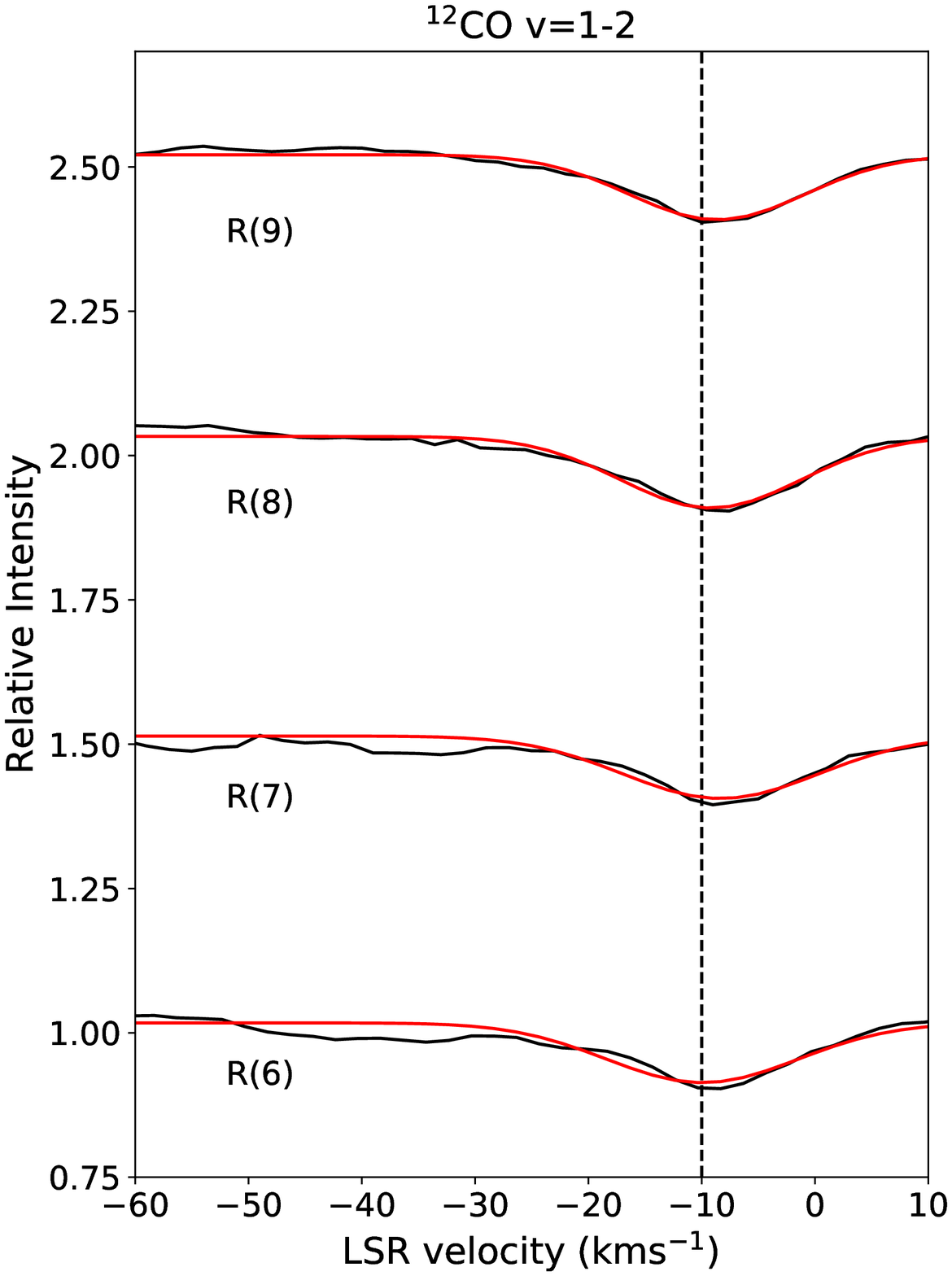}
\includegraphics[width=45mm,scale=1.5]{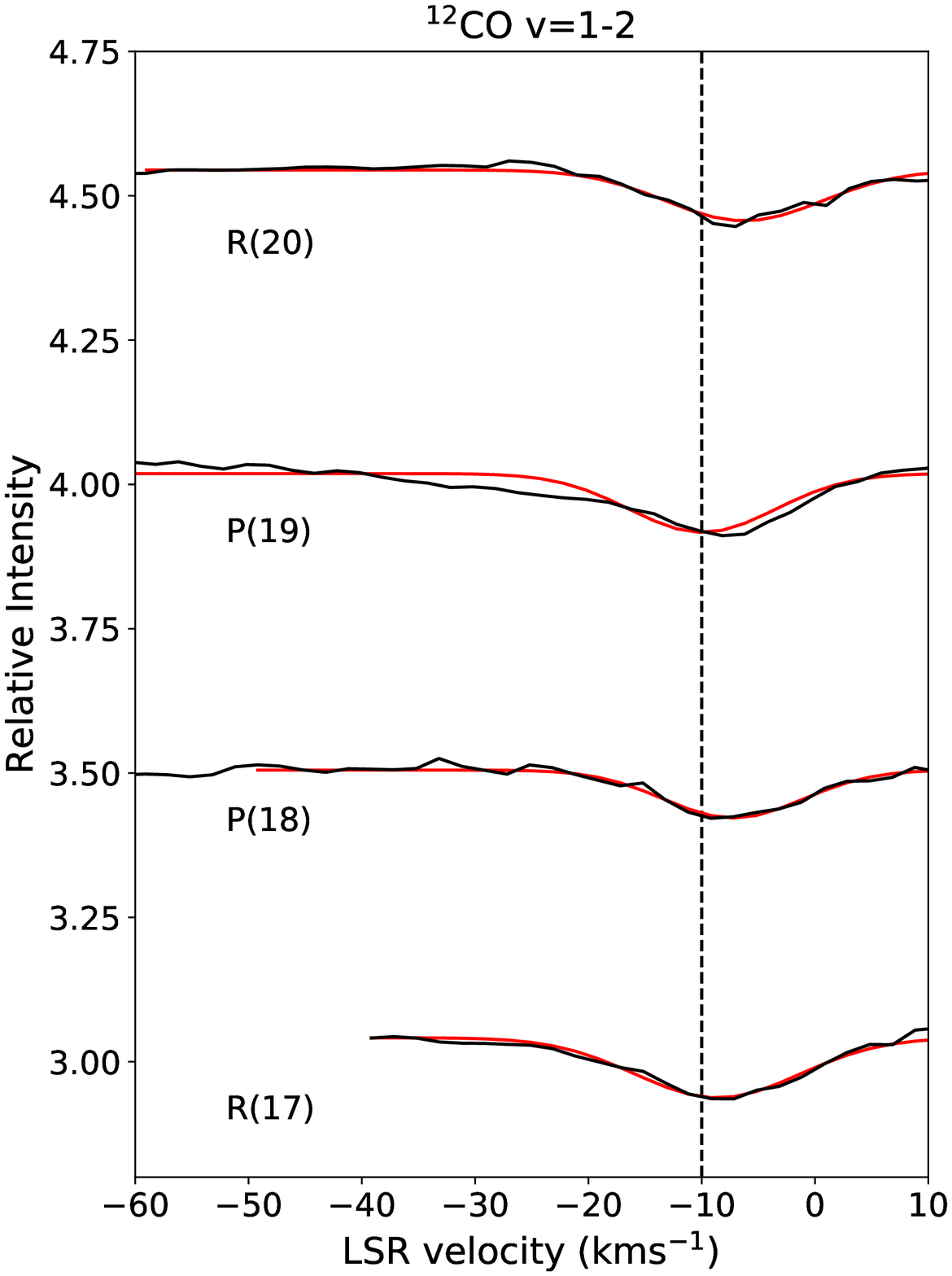}
\caption{All lines of CS, $^{13}$CO and $^{12}$CO v=1-2. The blue dotted lines represent the individual velocity components and the red solid lines show the overall fit. The black dashed line denotes -10 kms$^{-1}$.}
\label{linesall}
\end{figure}

\end{document}